\documentclass{article}            
\usepackage[dvips,final]{graphics}

\begin{document}                 

\title{The new very small angle neutron scattering spectrometer at Laboratoire L\'eon Brillouin}

\author{S. D\'esert$^{1}$ \and V. Th\'evenot$^{1}$ \and J. Oberdisse$^{2}$ \and A. Br\^ulet$^{1}$}

\maketitle                      

\begin{abstract}
The design and characteristics of the new very small angle neutron scattering spectrometer under construction at the Laboratoire L\'eon Brillouin is described.
Its goal is to extend the range of scattering vectors magnitudes towards 2$\times$10$^{-4}$ \AA$^{-1}$.
The unique feature of this new spectrometer is a high resolution two dimensional image plate detector sensitive to neutrons.  
The wavelength selection is achieved by a double reflection supermirror monochromator and the collimator uses a novel multibeam design.
\end{abstract}

\footnote{Laboratoire L\'eon Brillouin, CEA Saclay, 91191 Gif-Yvette, France}
\footnote{Laboratoire des Collo\"\i des, Verres et Nanomat\'eriaux, 34095 Montpellier, France}

\newpage

\section{Introduction}
More and more studies of large-scale objects ($>$ 50 nm) are needed nowadays. The very small angle neutron spectrometer, VSANS (Tr\`es Petits Angles, TPA), will be dedicated to the study of such nanostructures, usually found in aggregates, branched polymers, organized multi-component systems (such as vesicles), reinforced rubber, cell membranes in biology, clays, porous systems, alloys in metallurgy, vortex lattice in supraconductors.
\\TPA is developed to reach a scattering vector modulus, $q$, down to 2$\times$10$^{-4}$ \AA$^{-1}$ with $q=(4\pi / \lambda)\sin(\theta /2)$ where $\theta$ is the scattering angle and $\lambda$ is the neutron wavelength. Such a small $q$ value is lower than those obtained by classical small angle scattering pinhole spectrometers in order to fill the gap between light scattering and classical small angle neutron scattering (SANS). It is very different from the ultra small angle neutron scattering spectrometers which can reach scattering vectors still one order of magnitude lower by using double crystal monochromators \cite{Bonse65}. The principles of TPA are the same as for classical SANS spectrometers: pinhole collimation but with very small apertures and two dimensional detector to ensure measurements in a wide wave vector space. The scattering range is designed to vary from 2$\times$10$^{-4}$ to 10$^{-2}$ \AA$^{-1}$, either by changing the wavelength and/or the detector position. In order to achieve measurements in such a wide $q$ range and keep a reasonnable flux, the multibeam technique \cite{glinka86} is settled for the collimation. This technique consists in using a set of masks defining several beams converging onto the detector to increase the flux at the sample position. In this paper, we describe the early measurements achieved with a prototype made of 7 masks. Another originality of this new spectrometer is a two dimensional high resolution image plate detector sensitive to neutrons \cite{wilkinson92}. Its sensitivity to $\gamma$ radiation leads us to use a double reflection supermirror monochromator instead of a velocity selector commonly manufactured with strong $\gamma$ emitters. We also present the first measurements obtained with this equipment.

\section{Spectrometer design}
\begin{figure}
\caption{Scheme of the spectrometer TPA at LLB (units in mm, drawing not to scale).}
\includegraphics{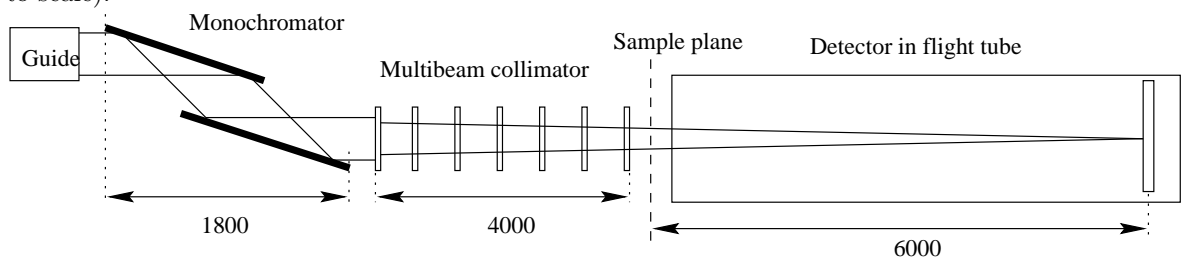}
\label{montage}
\end{figure}

TPA is installed at the end of a cold neutron bender G5bis whose inner dimensions of guide exit are 50 mm height and 25 mm width. The main constraint on the design of the instrument is the 12 m floor space available from the guide exit in the reactor guide hall. The layout of the instrument is shown in Figure \ref{montage}.

\subsection{Two-dimensional position sensitive detector}
\begin{figure}
\caption{Sample of latex-silica nanocomposite film at pH=5 \cite{oberdisse02} measured on D11 with a distance of 36.7 m and $\lambda$=10 \AA \hspace{0.1cm} ($\circ$) and TPA with a distance of 4 m and $\lambda$=7 \AA \hspace{0.1cm} ($\bullet$). Note the q-axis in log scale.}
\includegraphics{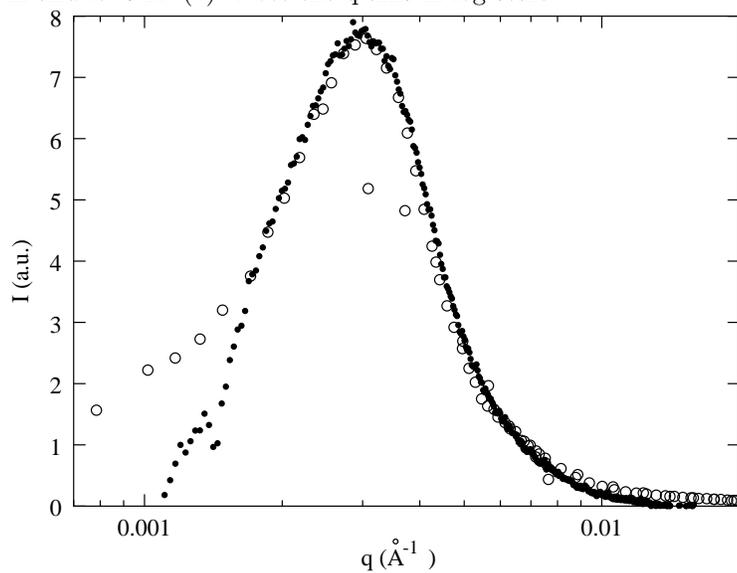}
\label{tpad11}
\end{figure}

The manufacture of a large multi-detector with small pixel size ($<$ mm) is important to achieve high resolution SANS measurements and avoids long length instrument, but is far from being easy. Therefore, it was decided to use a commercial image plate for X-ray (MAR345, Marresearch GmbH), equipped with a neutron converter (Gd$_2$O$_3$) and a storage imager (BaFBr doped with Eu$^{2+}$) read in-situ. Such a detector allows high sensitivity and dynamics \cite{zemb93}. Its dimensions are 2300 $\times$ 2300 pixels of 150 $\mu$m each. Such a definition is much higher than that of one among the best 2D SANS spectrometer, D11 at ILL. Figure \ref{tpad11} presents a latex-silica nanocomposite film \cite{oberdisse02} measured on D11 and TPA for demonstration (the deviation at low scattering wave vector is due to the lower q-min of D11 in this configuration).
The detector will lie in an hermetically sealed flight tube under Helium atmosphere and will move between 1 and 6 m far from the sample plane. The drawback of such detectors is their sensitivity to $\gamma$ radiations which imposes special care such as the use of lead and enriched lithium ($^6$Li) neutron absorber (instead of cadmium or gadolinium) as well as heavy concrete shielding around the detector.

\subsection{Wavelength selection}
\begin{figure}
\caption{Typical reflectivity curve for two supermirror monochromator 3$\theta_c$ measured on EROS \cite{menelle03} at LLB with an angle of 1.5$^{\circ}$.}\includegraphics{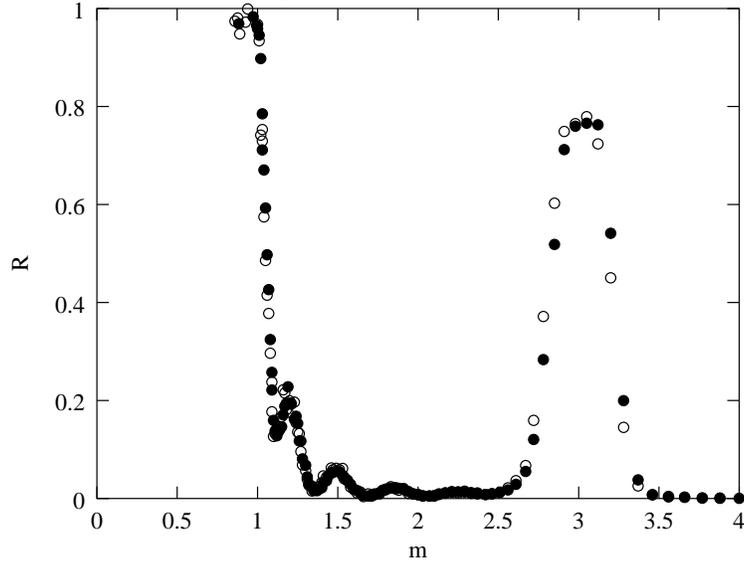}
\label{M2A}
\end{figure}

\begin{figure}
\caption{Time of flight measurements for neutrons of wavelength, 5.4 ($\circ$), 8 ($\bullet$) and 12 \AA ($\diamond$). The intensities have been rescaled to unity for convenience. Measurements were made with a chopper (5 mm slit and 23 cm radius) at a rotation speed of 3000 rpm and 2 mm diameter pinhole, the detector is located at 3.22 m from the chopper. The instrumental spreading is approximatively 0.24 \AA.}
\includegraphics{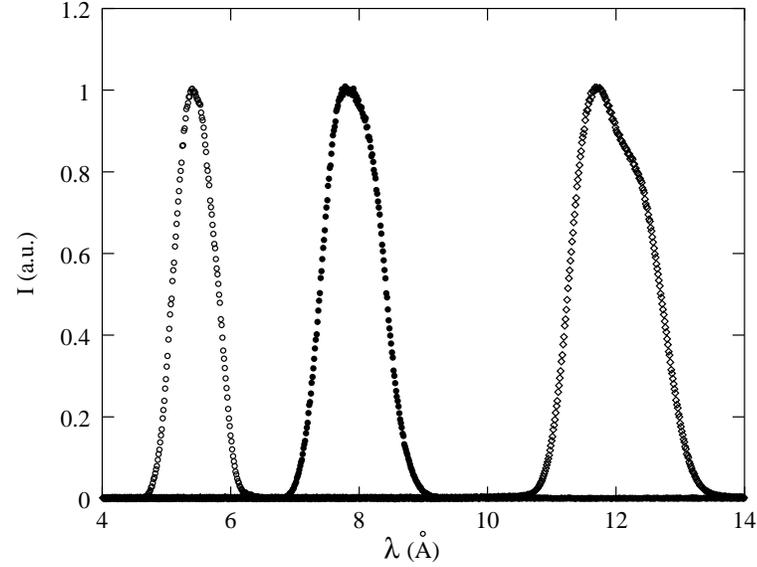}
\label{ToF}
\end{figure}

A conventional velocity selector can not be used because of the strong $\gamma$ radiation emitted from gadolinium which is not compatible with our detector sensitivity to $\gamma$.
Therefore a double reflection supermirror monochromator has been developed. It is made of two supermirrors (purchased at Swissneutronics and mounted by CILAS) with a critical angle 3$\alpha_{c}$ (3 times the critical angle of Ni) and a bandwidth around 15$\%$. Each mirror is 30 mm high and 60 cm long made from two pieces each of 30 cm long glued on a glass support. The characteristics of these supermirror monochromators are especially interesting for our purpose: 80\% transmission, no direct view of the guide and weaker $\gamma$ production. The total transmission after double reflection is therefore 64\% and remains constant for all wavelengths. Figure \ref{M2A} shows the reflection curves for both supermirrors as a function of $m$, the critical angle normalized to that of Ni: both supermirrors have a ratio $\frac{\Delta m}{m}$ of 0.14.\\ 
Wavelength selection is achieved by rotation of the mirrors relative to the beam axis. Indeed, if $\alpha$ represents the angle between the beam axis and the mirror plane, then the reflected wavelength will be given by:
\begin{equation}
\lambda = \frac{\alpha}{m\alpha_{c}}
\end{equation}
with $m$=3 and $\alpha_{c}$, the critical angle of Ni, around 0.1$^{\circ}$\AA$^{-1}$.\\
The two mirrors are mounted on a rotation stage and the second one is also mounted on a 0.8 m length translation stage parallel to the beam axis in order to keep fixed the beam axis after the monochromator (with 55 mm offset relative to the original axis). This 1.8 m long monochromator is under vacuum. With this setup, the chosen wavelength can continuously be varied from 5 up to 20 \AA \hspace{0.1cm} corresponding to mirrors rotations of 1.5 up to 6$^{\circ}$.
Figure \ref{ToF} presents time of flight measurements for three different monochromator configurations. For each wavelength, $\frac{\Delta \lambda}{\lambda}$ equals 0.11 and is consistent with the product of the reflection curves of the two supermirrors from figure \ref{M2A}. We observe a shoulder in the peak for neutrons of wavelength 12 \AA \hspace{0.1cm} and until now we have no clear-cut explanation concerning this effect. Anyway, it has negligible effect on the final resolution of the incident monochromated beam.
 
\subsection{Multibeam collimation}
\begin{figure}
\caption{Representation of the pinhole geometry (units in mm, drawing not at scale) of the mask at the entrance of the collimator, the pinhole diameter is 1.6 mm.}
\includegraphics{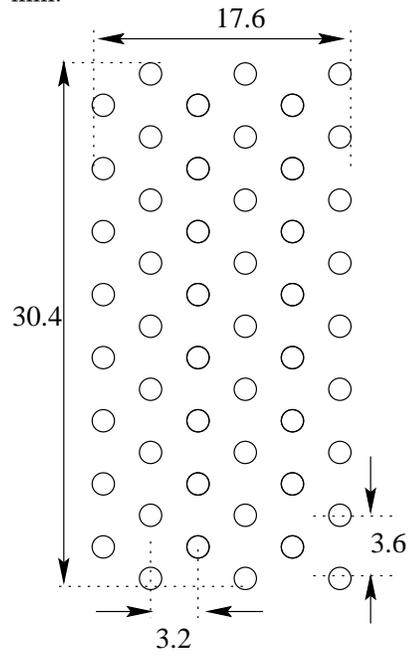}
\label{masque51}
\end{figure}

\begin{figure}
\caption{Comparison of the transmitted beam intensity measured by the detector with simple ($\bullet$) and 16 multi beam ($\circ$) collimation. The entrance and exit pinhole diameters are 1.6 and 1.0 mm, respectively and the wavelength is 7 \AA. The intensity gain is 12.}
\includegraphics{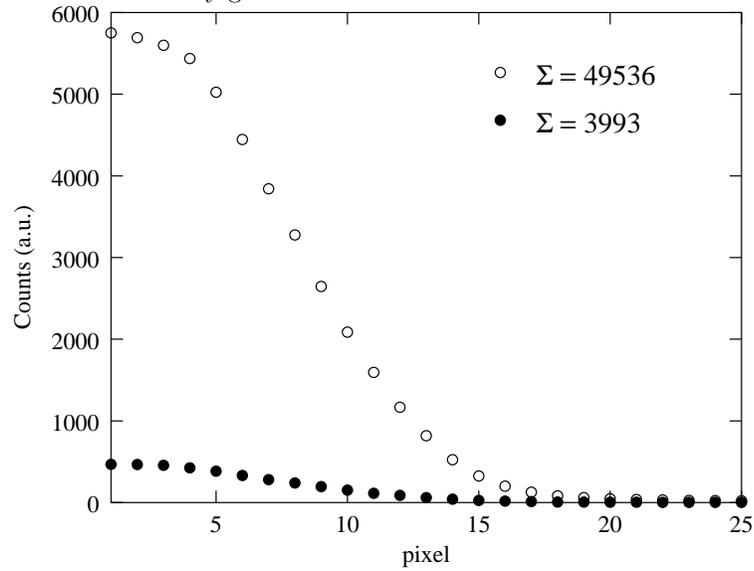}
\label{FV16}
\end{figure}

The weak neutron flux on the detector area, due to the tiny collimation, the small pixel size and the inherent flux for large wavelengths is the main limitation of this kind of spectrometer. Among the available focusing techniques, we have chosen the multibeam technique rather than a set of lenses or a focusing mirror. The use of lenses improves the intensity and gives access to lower minimum scattering vectors \cite{choi00} but also have drawbacks such as requiring to change the number of lenses when changing the wavelength. It also adds SANS scattering from the lens material and parasitic diffusion due to surface roughness or manufacturing imperfections and induces chromatic abberation. In the case of magnetic lenses \cite{oku04}, 50\% of the flux is already lost due to polarization. We also didn't consider beam focusing with mirror \cite{alefeld97} because of the large sample volumes required. Therefore, we are developing a multibeam converging collimator in order to recover reasonable incident flux. A multibeam prototype collimator of 4 m length was successfully tested. It features 7 masks of 51 holes each (Figure \ref{masque51}) acting as 51 beams converging on the detector plane. The neutron gain should be roughly equal to the number of beams. Each mask is made from $^6$Li oxide powder mixed in an epoxy matrix. Their outer dimensions are 36 $\times$ 50 mm$^2$ and 4 mm thickness. The pinholes have 1.6 and 1.0 mm diameters at entrance and exit of collimator, respectively.
They are distributed on an hexagonal array in order to maximize their number on a given area. The distance between two consecutive pinhole centres for the entrance mask is 3.6 mm vertically and 3.2 mm horizontally. This leads to a gap of 2 mm between the edges of two consecutive pinholes. Since 0.5 mm gap of matter between two neighbouring holes is close to the masks manufacturing limit, this value will be used for the exit mask of the future collimator. Coordinates of the holes of the following masks are deduced by homothety. The number of masks required to avoid crosstalk (neutron passing through the set of masks but not focusing onto the detector) only depends on the distance between the holes: the smaller the distance between pinholes, the higher the number of masks required. We used acceptance angle tracing \cite{copley90} as well as Vitess \cite{zsigmond02} and McStas \cite{Lefman99} Monte Carlo simulations to show that 7 masks, equally spaced, would absorb all unwanted neutrons.
Figure \ref{FV16} shows the intensities obtained with the transmitted beam using the multi beam prototype with 16 pinholes and 1 pinhole (15 pinholes masked). The ratio between the measured intensities is 12, thus close to the expected ratio of 16 assuming the beam is homogeneous and the pinholes perfectly manufactured and aligned.
The masks are mounted on a vertical stage to compensate the neutron fall due to gravity. Indeed, a neutron of wavelength 20 \AA \hspace{0.1cm} falls 2 mm after a 4 m flight path and this value is comparable to the exit pinhole diameter of 1 mm.
The masks are also mounted on horizontal stages acting as masks changer when the detector position will change.
The final version of the collimator will be composed of 3 sets of masks to focus the beam at 3 favoured detector positions: 1, 3 and 6 m. A fourth set of masks could also be implemented with multi slits along the vertical axis \cite{barker06}, 1.6 mm width at the collimator entrance, optimized for the 6 m detector position. These multi slits would be used in the case of weakly and isotropic scattering samples for an improved theoretical gain of a factor of 60 compared to multi pinholes. Although the slit smeared intensity will undergo a different log-log slope, suitable programs allow for correct data treatment \cite{glatter77}.

\section{Conclusion}
The new VSANS spectrometer, TPA, already works in a non definitive version. Its final version should give access to scattering vector magnitudes not accessible by standard SANS spectrometers, typically from 2$\times$10$^{-4}$ to 10$^{-2}$ \AA$^{-1}$. It features a newly designed double reflection supermirror monochromator, a prototype multibeam collimator and an image plate detector.
\\
\textbf{Aknowledgment}\\
The authors wish to thank P. Permingeat and A. Gabriel for the designs and for useful discussions, F. Coneggo and P. Lambert for the electronic devices and A. Menelle for the reflectivity measurements.\\
This research project has been supported by the European Commission under the 6th Framework Programme through the Key Action: Strengthening the European Research Area, Research Infrastructures. Contract n°: RII3-CT-2003-505925.

\bibliographystyle{unsrt}
\bibliography{iucr_sas2006}

\begin{thebibliography}{10}

\bibitem{Bonse65}
U.~Bonse and M.~Hart.
\newblock unknown.
\newblock {\em Appl. Phys. Lett.}, 7:238--240, 1965.

\bibitem{glinka86}
C.J. Glinka, J.M. Rowe, and J.G. LaRock.
\newblock unknown.
\newblock {\em J. Appl. Cryst.}, 19:427--429, 1986.

\bibitem{wilkinson92}
C.~Wilkinson, A.~Gabriel, M.S. Lehmann, T.~Zemb, and F.~N\'e.
\newblock unknown.
\newblock {\em SPIE}, 1737:324--329, 1992.

\bibitem{zemb93}
F.~N\'e, D.~Gazeau, J.~Lambard, P.~Lesieur, T.~Zemb, and A.~Gabriel.
\newblock unknown.
\newblock {\em J. Appl. Cryst.}, 26:763--773, 1993.

\bibitem{oberdisse02}
J.~Oberdisse and B.~Dem\'e.
\newblock unknown.
\newblock {\em Macromolecules}, 35:4397--4405, 2002.

\bibitem{choi00}
S.-M Choi, J.G. Barker, C.J. Glinka, Y.T. Chang, and P.L. Gammel.
\newblock unknown.
\newblock {\em J. Appl. Cryst.}, 33:793--796, 2000.

\bibitem{oku04}
T.~Oku, J.~Suzuki, H.~Sasao, T.~Adachi, T.~Shinohara, K.~Ikeda, T.~Morishima,
  K.~Sakai, Y.~Kiyanagi, M.~Furusuka, and H.M. Shimizu.
\newblock unknown.
\newblock {\em Nucl. Instr. and Meth. A}, 529:116--119, 2004.

\bibitem{alefeld97}
B.~Alefeld, Hayes C., F.~Mezei, D.~Richter, and T.~Springer.
\newblock unknown.
\newblock {\em Physica B}, 234-236:1052--1054, 1997.

\bibitem{copley90}
J.R.D. Copley.
\newblock unknown.
\newblock {\em Nucl. Instr. Methods A}, 287:363--373, 1990.

\bibitem{zsigmond02}
D.~Zsigmond, K.~Lieutenant, and F.~Mezei.
\newblock unknown.
\newblock {\em Neutron News}, 13(4):11--14, 2002.

\bibitem{Lefman99}
K.~Lefman and K.~Nielsen.
\newblock unknown.
\newblock {\em Neutron News}, 10:20--23, 1999.

\bibitem{barker06}
J.~Barker.
\newblock Performance review of various advanced optical designs for very
  small-angle neutron scattering (vsans) instrumentation.
\newblock Report, NIST Center for Neutron Research, National Institue of
  Standards and Technology, Gaithersburg, USA, 2006.

\bibitem{glatter77}
O.~Glatter.
\newblock unknown.
\newblock {\em J. Appl. Cryst.}, 10:415--421, 1977.

\bibitem{menelle03}
A.~Menelle, J.~Jestin, and F.~Cousin.
\newblock unknown.
\newblock {\em Neutron News}, 14(3):26--30, 2003.

\end{thebibliography}
\newpage

\end{document}